\documentclass{article}
\def\ba{\begin{array}}
\def\ea{\end{array}}
\def\be{\begin{equation}}
\def\ee{\end{equation}}

\def\a{{\alpha}}

\def\bea{\begin{eqnarray}}
\def\eea{\end{eqnarray}}
\def\b{\beta}

\def\L{\Lambda}

\def\b{\beta}

\begin{document}
\begin{titlepage}

\hfill{}
\vskip 5 mm
\noindent{ \Large \bf Phase transition in an asymmetric generalization of 
           the zero--temperature Glauber model}

\vskip 1 cm
\noindent{Mohammad Khorrami$^{1,a}$ \& Amir Aghamohammadi$^{2,b}$}
\vskip 5 mm
{\it
  \noindent{ $^1$ Institute for Advanced Studies in Basic Sciences,
             P.O.Box 159, Zanjan 45195, Iran. }

  \noindent{ $^2$ Department of Physics, Alzahra University,
             Tehran 19834, Iran. }

  \noindent{ $^a$ mamwad@iasbs.ac.ir}

  \noindent{ $^b$ mohamadi@theory.ipm.ac.ir}
  }
\vskip 1 cm



\vskip 1cm

\begin{abstract}
An asymmetric generalization of the zero--temperature Glauber
model on a lattice is introduced. The dynamics of the particle--density
and specially the large--time behavior of the system is studied. It is
shown that the system exhibits two kinds of phase transition, a static
one and a dynamic one.
\end{abstract}
\end{titlepage}
\section {Introduction}
In recent years, reaction--diffusion systems have been studied by many
people. As mean--field techniques, generally, do not give correct results
for low--dimensional systems, people are motivated to study
exactly--solvable stochastic models in low dimensions.
Moreover, solving one--dimensional systems should in principle
be easier. Exact results for some models on a one--dimensional lattice
have been obtained, for example in [1--12].
Different methods have been used to study these models, including
analytical and asymptotic methods, mean field methods, and large--scale
numerical methods.

Some interesting problems in non--equilibrium systems are
non--equilibrium phase transitions described by
phenomenological rate equations, and the way the system relaxes to
its steady state. Kinetic generalizations of the Ising model, for example
the Glauber model or the Kawasaki model, are such phenomenological models
and have been studied extensively [13--18]. Combination of the Glauber and
the Kawasaki dynamics has been also considered [19--21].

In this article, we want to study an asymmetric generalization of the
zero--temperature Glauber model on a lattice with boundaries.
There are also
sources (or sinks) of particles at the end points of the lattice.
We study the dynamics of the particle density, and specially
the large time behavior of the system. In the thermodynamic limit,
the system shows two kinds of phase transition. One of these is a
static phase transition, the other a dynamic one.
The static phase transition is controlled by the reaction rates, and
is a discontinuous change of the behavior of the derivative of the
stationary particle density at the end points, with respect to the
reaction rates. The dynamic phase transition is controlled by
the injection- and extraction- rates of the particles at the end points,
and is a discontinuous change of the relaxation time towards the
stationary configuration.

\section{Asymmetric Glauber model at zero temperature}
In the ordinary Glauber model, the interaction is between three
neighboring sites.
Spin flip brings the system to equilibrium with a heat bath at
temperature $T$.
A spin is flipped with the rate $\mu := 1 - \tanh{ J\over kT}$
if the spin of both of its neighboring sites are the same as itself,
and is flipped with the rate $\lambda := 1 + \tanh{ J\over kT}$
if the spin of both of its neighboring sites are opposite to it.
At domain boundaries, the spins are flipped with unit rate.
So the interactions can be written as,
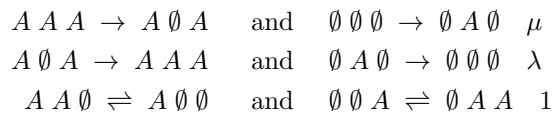
\begin{eqnarray}
A\; A \; A \; \to \; A\; \emptyset \; A \; &\mbox{ and }&
\emptyset \; \emptyset \; \emptyset \; \to \; \emptyset \; A
\; \emptyset \quad \mu \nonumber \\
A \; \emptyset \; A \; \to \; A \; A \; A \; &\mbox{ and }&
\emptyset \; A \; \emptyset  \; \to
 \; \emptyset \; \emptyset \; \emptyset \quad \lambda \nonumber \\
A \; A \; \emptyset \; \rightleftharpoons \;
A \; \emptyset  \; \emptyset \; &\mbox{ and }&
\emptyset  \; \emptyset \; A \;
\rightleftharpoons \; \emptyset \; A \; A\quad 1 \nonumber
\end{eqnarray}
where spin up and spin down are denoted by $A$ and $\emptyset$.
One can interpret an up spin as a  particle, and a down spin as
a hole. At zero temperature, the Glauber dynamics is effectively a
two--site interaction \cite{GS}:
\be
A\emptyset   \to \cases{A A \cr
\emptyset \emptyset  \cr}
\ee
\be
\emptyset  A  \to \cases{A A \cr
\emptyset \emptyset  \cr},
\ee
where all the above processes occur with  the same rate.

One can consider the following interactions,
as an asymmetric generalization of the zero--temperature Glauber model.
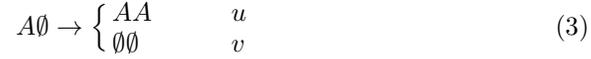
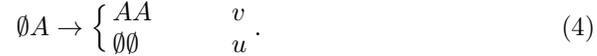
\be
A\emptyset   \to \cases{A A & $\qquad u $\cr
\emptyset \emptyset  & $\qquad v$\cr}
\ee
\be
\emptyset  A  \to \cases{A A & $\qquad v $\cr
\emptyset \emptyset  & $\qquad u $\cr}.
\ee
If $u\ne v$, the above system has left--right asymmetry.
The above system on an infinite lattice has been investigated in
\cite{AM}, where its $n$--point functions, its equilibrium states,
and its relaxation towards these states are studied. It can
be easily shown that the time evolution equation for the average densities
of the system with the above interactions are the same as that of a system
with the  following interactions,
where diffusion is also present:
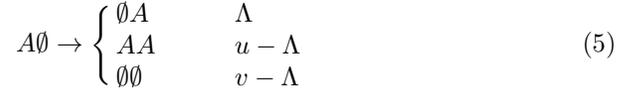
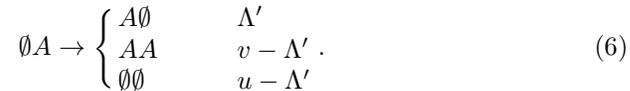
\be
A\emptyset   \to \cases{\emptyset  A & $\qquad \L$ \cr
A A & $\qquad u - \L$\cr
\emptyset \emptyset  & $\qquad v- \L$\cr}
\ee
\be
\emptyset  A  \to \cases{A\emptyset   & $\qquad \L'$ \cr
A A & $\qquad v - \L'$\cr
\emptyset \emptyset  & $\qquad u- \L'$\cr}.
\ee

Consider a lattice with $L$ sites and an asymmetric zero--temperature
Glauber dynamics as the interaction. The rates of injection
and extraction of
particles in the first site (final site) are $a$ and $a'$ ($b$ and $b'$),
respectively.
The time evolution equations for the average densities are then
\bea \label{nk}
\langle \dot n_k \rangle& =& -(u+v)\langle  n_k\rangle +u
\langle  n_{k-1}\rangle + v\langle  n_{k+1}\rangle \qquad
{\rm for} \quad k\ne 1,L \cr
\langle \dot n_1 \rangle& =&a-(a+a'+v)\langle  n_1\rangle +
 v\langle  n_2\rangle \cr
\langle \dot n_L \rangle &=&b-(b+b'+u)\langle  n_L\rangle +
 u\langle  n_{L-1}\rangle .
\eea
First, let us calculate the profile of average densities at large
times. At large times, the system goes to its stationary state, and the
time--derivatives of the left--hand sides vanish. One can see then that the
solution to the above system is
\be
\langle n_k (\infty )\rangle =\a +\b \left({u\over v}\right)^k.
\ee
Putting this in (\ref{nk}), one can easily find $\a$ and $\b$. In the
thermodynamic limit ($L\to\infty$), the solution becomes
\bea\label{001}
\langle n_k (\infty )\rangle&=&{a\over{a+a'}}+
{ba'-ab'\over (a+a')(b+b'+u-v)}\left( {u\over v}\right)^{k-L},\qquad u>v\cr
\langle n_k (\infty )\rangle&=&{b\over{b+b'}}+
{ab'-a'b\over (b+b')(a+a'+v-u)}\left( {u\over v}\right)^{k-1},\qquad u<v.
\eea
It is seen that the density profile is flat at the left end ($k\ll L$) for
$u>v$ and its value is independent of the reaction rates. But as $v$ exceeds
$u$, the density profile acquires a finite slope, proportional to
$\ln (u/v)$. This is the static phase transition previously mentioned.

Now return to the dynamics of the system. Let us write the homogeneous part
of (\ref{nk}) as
\be\label{a2}
\langle\dot n_k\rangle =h_k^l\langle n_l\rangle,
\ee
and find the eigenvalues and eigenvectors of the operator $h$. One finds
\bea\label{a3}
E x_k&=&-(u+v)x_k +u x_{k-1}+v x_{k+1},\qquad k\ne 1,L\cr
E x_1&=& -(a+a'+v)x_1+v x_2,\cr
E x_L&=& -(b+b'+u)x_L+u x_{L-1},
\eea
where the eigenvalue and eigenvector have been denoted with $E$ and $x$,
respectively. The solution to these equations is
\be\label{a4}
x_k=\alpha z_1^k+\beta z_2^k,
\ee
where $z_i$'s satisfy
\be\label{a5}
E=-(u+v)+vz+{u\over z},
\ee
and
\bea\label{a6}
v(\alpha z_1^2+\beta z_2^2)-(E+a+a'+v)(\alpha z_1+\beta z_2)&=&0\cr
u(\alpha z_1^{L-1}+\beta z_2^{L-1})-(E+b+b'+u)(\alpha z_1^L+\beta z_2^L)&=&0.
\eea
To have nonzero solutions for $x$, these last two equations should be
dependent, the criterion for which is
\be\label{a7}
(u+z_1\delta a)(v z_2^{L+1}+z_2^L\delta b)-(u+z_2\delta a)(v z_1^{L+1}+
z_1^L\delta b)=0,
\ee
where (\ref{a5}) has been used to eliminate $E$, and $\delta a:=a+a'-u$ and
$\delta b:=b+b'-v$.
Defining
\bea\label{a9}
Z_i&:=&z_i\sqrt{v\over u}\cr
A&:=&{{\delta a}\over\sqrt{uv}}\cr
B&:=&{{\delta b}\over\sqrt{uv}},
\eea
(\ref{a7}) is simplified to
\be\label{a10}
Z^{-(L+1)}(1+AZ)(1+BZ)-Z^{L+1}(1+AZ^{-1})(1+BZ^{-1})=0.
\ee
The eigenvalue $E$ satisfies
\be\label{a11}
E=-(u+v)+\sqrt{uv}(Z+Z^{-1}).
\ee
Two obvious solutions of the equation (\ref{a10}) are $Z=\pm 1$. But these
generally don't correspond to eigenvalues and eigenvectors. In fact for
these solutions, $Z$ and $Z^{-1}$ are the same, so that (\ref{a4}) should
be modified to
\be\label{12a}
x_k=(\alpha +\beta k)(\pm 1)^k,
\ee
and it is not difficult to see that these do not fulfill the boundary
conditions unless $\alpha =\beta =0$. Equation (\ref{a11}) can be written
as a polynomial equation of order $2(L+1)$, and hence has $2L$ more roots
in addition to $Z=\pm 1$. For these $2L$ roots, if $Z$ is a root $Z^{-1}$ is
another root, and these two correspond to one eigenvalue and one
eigenvector. So the $L\times L$ matrix $h$ does have $L$ eigenvalues and
eigenvectors.

For $A=B=0$, (\ref{a10}) is very simple and its nontrivial solutions are
\be\label{a12}
Z_{(s)}=e^{i\pi s/(L+1)}, \qquad 1\leq s\leq L,
\ee
and their inverses.
All of these are phases and the real--part of the corresponding eigenvalues
satisfy
\be\label{a13}
{\rm Re}(E)\leq -(u+v)+2\sqrt{uv}\cos \left({\pi\over{L+1}}\right)<0.
\ee
The maximum of the real--part of the eigenvalues determines the relaxation
time toward the stationary average--density profile. That is
\be\label{a14}
\tau=\left[u+v-2\sqrt{uv}\cos \left({\pi\over{L+1}}\right)\right]^{-1}.
\ee
In the limit $L\to\infty$, this is simplified to
\be\label{a15}
\tau=(u+v-2\sqrt{uv})^{-1}.
\ee
The general solution to (\ref{nk}) is seen to be
\be
\langle n_k (t)\rangle =\sum_{s,m} {2\over L+1} \exp[E_{(s)} t]
\langle n_k (0)\rangle \left({u\over v}\right)^{(k-m)/2}
\sin\left({\pi sk\over L+1}\right)\sin\left({\pi sm\over L+1}\right).
\ee
Now consider the general case. The equation (\ref{a10}) can be written as
\be\label{a17}
G(Z):=F(Z)-F(Z^{-1})=0,
\ee
where
\be\label{a18}
F(Z):=Z^{-(L+1)}(1+AZ)(1+BZ).
\ee
If $Z$ is a phase, it satisfies (\ref{a10}) provided $F(Z)$ is real.
Consider the phase of $F(Z)$ for unimodular $Z$. We have
\be\label{a19}
\phi[F(z)]=-(L+1)\phi(Z)+\phi(1+AZ)+\phi(1+BZ),
\ee
where $\phi$ denotes the phase of its argument. As the phase of $Z$ is
changed from $0$ to $2\pi$, the change of the phase of $F(z)$ is
\be\label{a20}
\Delta\phi[F(Z)]=-2\pi(L+1)+\Delta\phi(1+AZ)+\Delta\phi(1+BZ).
\ee
For $|A|,|B|<1$, the phase changes of $(1+AZ)$ and $\phi(1+BZ)$ is zero, as
$Z$ moves on the whole unit circle. So,
\be\label{a21}
\Delta\phi[F(Z)]=-2\pi(L+1), \qquad \hbox{for} |A|,|B|<1.
\ee
But this means that the phase of $F(Z)$ will be an integer multiple of $\pi$
for at least $2(L+1)$ points on the unit circle. So all of the solutions
of (\ref{a10}) are still phases, although they may be not uniformly spaced
on the unit circle. Two of these are $Z=\pm 1$. The remaining $2L$ points
correspond to $L$ eigenvalues and eigenvectors for $h$. One concludes that
for $|A|,|B|<1$,
\be\label{a22}
\tau=(u+v-2\sqrt{uv}\cos\theta)^{-1},
\ee
for some $\theta$. Specially, at $L\to\infty$, the relaxation time is the
same as the relaxation time for $A=B=0$, that is the same as (\ref{a15}).

If, for example, $|A|>1$, then the total phase change of $(1+AZ)$ is no
longer zero. It is $2\pi$. One may then lose two of the roots of the unit
circle. Note that the mere fact that $\Delta\phi[F(Z)]=-2\pi L$ does not
mean that there are just $2L$ solutions of (\ref{a10}) on the unit circle,
since the phase of $F(Z)$ needn't be monotonic. To find values of $A$ and
$B$, for them the number of the solutions of (\ref{a10}) on the unit circle
is $2L$ or $2(L-1)$, consider the function $G(Z)$ at the points $Z=\pm 1$.
Increasing $|A|$ or $|B|$, two of the roots on the unit circle tend to $1$
or $-1$, and then move out of the unit circle and on the real line. At the
point that this occurs, either $G'(1)$ or $G'(-1)$ become zero, as there
will be multiple roots at $\pm 1$. So the criterion for each change (losing
$2$ roots of the unit circle) is that either $G'(1)$ or $G'(-1)$ become
zero. The curves in the $AB$ plane, corresponding to these changes are
\be\label{a23}
L(1+A)(1+B)+1-AB=0,\qquad G'(1)=0,
\ee
and
\be\label{a24}
L(1-A)(1-B)+1-AB=0,\qquad G'(-1)=0.
\ee
These curves divide the plane into six regions:
\begin{itemize}
\item[I:] all of the solutions are phases.
\item[II:] $2L$ phase solutions, 2 real negative solutions.
\item[III:] $2(L-1)$ phase solutions, 4 real negative solutions.
\item[IV:] $2L$ phase solutions, 2 real positive solutions.
\item[V:] $2(L-1)$ phase solutions, 4 real positive solutions.
\item[VI:] $2(L-1)$ phase solutions, 2 real negative solutions, 2 real
positive solutions.
\end{itemize}
In the above, by real solutions it is meant real solutions besides the
trivial solutions $\pm 1$.
Note, however, that not all of this plane is physical. The physical region
is that part of the plane, which corresponds to nonnegative values for
the injection and extraction rates. Returning to the
definitions of $A$ and $B$, equation (\ref{a9}), it is seen that
\bea\label{a25}
A_{\rm{min}}&=&-\sqrt{u\over v}\cr
B_{\rm{min}}&=&-\sqrt{v\over u}.
\eea
The point $(A_{\rm{min}}, B_{\rm{min}})$ itself is on the curve
\be
A_{\rm{min}}B_{\rm{min}}=1.
\ee
One can see that, unless $u=v$, part of the physical region is
in the region IV, where two of the solutions of (\ref{a10}) are real
positive (and of course inverse of each other). This makes the maximum of
the real part of $E$ larger than $-(u+v)+2\sqrt{uv}$, and correspondingly the
relaxation time larger than (\ref{a15}).

In the thermodynamic limit $L\to\infty$, the regions of the $AB$ plane are
greatly simplified. In fact the curves corresponding to $G'(\pm 1)=0$ become
\be\label{a26}
A=-1,\quad\hbox{or}\quad B=-1,\qquad G'(1)=0,
\ee
and
\be\label{a27}
A=1,\quad\hbox{or}\quad B=1,\qquad G'(-1)=0.
\ee
In the case either $|A|$ or $|B|$ are greater than 1, the real roots of
(\ref{a10}) are simply
\be\label{a28}
Z=-A, -A^{-1},\qquad |A|>1
\ee
and
\be\label{a29}
Z=-B, -B^{-1},\qquad |B|>1
\ee
Then, if for example $A$ is negative and less than $-1$ and $A<B$, the
maximum real part of $E$ is $-(u+v)-\sqrt{uv}(A+A^{-1})$, and the relaxation
time of the system is
\be\label{a30}
\tau =[u+v+\sqrt{uv}(A+A^{-1})]^{-1},
\ee
which is greater than (\ref{a15}). This is a phase transition which occurs
at $A=-1$. For $A>-1$, the relaxation time is constant, (\ref{a15}). For
$A<-1$, it is $A$--dependent. The minimum of $A$ is $-\sqrt{u/v}$, for which
one of the eigenvalues of $h$ become zero, and the relaxation time becomes
infinite. The same effect is seen for $B<-1$. As mentioned before, for
$u=v$ (the ordinary zero--temperature Glauber model) no part of the region
IV is in the physical region, and this
transition does not occur. The phase transition discussed here, is the
dynamical phase transition mentioned before.

\vskip 2\baselineskip
\noindent{\bf Acknowledgement}\\
The authors would like to thank Institute for Studies in Theoretical 
Physics and Mathematics for partial support.
\newpage

\end{document}